\documentstyle[twoside,fleqn,espcrc2,epsfig]{article}
\newcommand{\be}{\begin{equation}}
\newcommand{\ee}{\end{equation}}
\newcommand{\ApJ}{{\it Astrophys. J.\,}}
\newcommand{\Nature}{{\it Nature\,}}
\newcommand{\NP}{{\it Nucl. Phys.\,}}
\newcommand{\PRL}{{\it Phys. Rev. Lett.\,}}

\begin{document}

\title{{\small{\hfill $\begin{array}{r} \mbox{DSF 38/2000, astro-ph/0012291}
\end{array}$}}\\
Primordial Nucleosynthesis, Cosmic Microwave Background and Neutrinos}

\author{G. Mangano\address{INFN, Sezione di Napoli, and Dipartimento di
Scienze Fisiche,\\Universit\`a di Napoli {\it Federico II}, Napoli, Italy},
A. Melchiorri\address{Dipartimento di Fisica, Universit\`a di Roma {\it La
Sapienza}, Roma, Italy, and\\Nuclear and Astrophysics LAb, University of
Oxford, Oxford, OX 3RH, UK}, and O. Pisanti\address{Dipartimento di Scienze
Fisiche, Universit\`a di Napoli {\it Federico II}, Napoli, Italy}}

\begin{abstract}
We report the results of a recent likelihood analysis combining the
primordial nucleosynthesis and the BOOMERanG and MAXIMA-1 data on cosmic
microwave background radiation anisotropies. We discuss the possible
implications for relic neutrino background of a high value for the baryonic
matter content of the universe, larger than what is expected in a standard
nucleosynthesis scenario.
\vspace{1pc}
\end{abstract}

\maketitle

In this contribution we report on a combined analysis of the dependence of
CMBR anisotropies and BBN on the energy fractions $\Omega_Bh^2$ and
$\Omega_\nu h^2 = N_{\nu} \Omega_\nu^0 h^2$ for {\it massless} neutrinos
($N_\nu$ standing for the effective neutrino number and $\Omega_\nu^0 h^2$
for the energy contribution of a single $\nu-\overline{\nu}$ specie). This
is aimed to test the standard and degenerate BBN scenario, using the recent
results of the BOOMERanG \cite{Boomerang} and MAXIMA-1 \cite{Maxima} CMBR
experiments and the measurements of $^4He$, $D$ and $^7Li$ primordial
abundances.

The theoretical tools necessary to achieve this goal are nowadays rather
robust. A new generation of BBN codes have been developed
\cite{Emmp1,Emmp2,LopTur}, which give the $^4He$ mass fraction $Y_p$ with a
theoretical error of few per mille, this effort being justified in view of
the small statistical error which is now quoted in the $Y_p$ measurements.
On the other hand, the theoretical predictions on the CMBR anisotropies
angular power spectrum have also recently reached a $1 \%$ level accuracy.
An important new insight is represented by the recent results obtained by
the BOOMERanG Collaboration \cite{Boomerang}. For the first time, in fact,
multifrequency maps of the microwave background anisotropies were realized
over a significant part of the sky, with $\sim 10'$ resolution and high
signal to noise ratio. In the last few years many results have been
obtained on light element primordial abundances as well \cite{dati}. The
$^4He$ mass fraction $Y_p$, has been measured with a $0.1 \%$ precision in
two independent surveys, from regression to zero metallicity in Blue Compact
Galaxies, giving a {\it low} value $Y_p^{(l)} = 0.234 {\pm} 0.003$, and a
high one $Y_p^{(h)} = 0.244 {\pm} 0.002$, which are compatible at $2\sigma$
level only, may be due to large systematic errors. In our analysis
\cite{Emmmp} we adopted the more conservative value $Y_p = 0.238 {\pm}
0.005$. A similar controversy holds in $D$ measurements, where observations
in different Quasars Absorption line Systems (QAS) lead to the incompatible
results $Y_D^{(l)} = \left(3.4 {\pm} 0.3 \right) 10^{-5}$, and $Y_D^{(h)} =
\left(2.0 {\pm} 0.5 \right) 10^{-4}$. We performed our analysis for both
low and high $D$ data. Finally, the most recent estimate for $^7Li$
primordial abundance, from the {\it Spite plateau} observed in the halo of
POP II stars, gives $Y_{^7Li} =\left(1.73 {\pm} 0.21 \right) 10^{-10}$. The
light nuclide yields strongly depend on the baryon matter content of the
universe, $\Omega_B h^2$. In particular, assuming a standard BBN scenario,
i.e. vanishing neutrino chemical potentials, the likelihood analysis gives,
at $95 \%$ C.L. \cite{Emmp2},
\be
\begin{array}{lcc} \mbox{low D} &
\Omega_B h^2 = .017 {\pm} .003 & 1.7 \leq N_\nu \leq 3.3, \\
\mbox{high D} &
\Omega_B h^2 = .007^{+.007}_{-.002}&  2.3 \leq N_\nu \leq 4.4.
\end{array}
\label{standard}
\ee

Regarding CMBR data, the anisotropy power spectrum, $C_\ell$, was measured
in a wide range of angular scales from multipole $\ell \sim 50$ up to $\ell
\sim 600$, with error bars of the order of $10 \%$, showing a peak at
$\ell_{peak}=(197 {\pm} 6)$ with an amplitude $DT_{200}=(69 {\pm} 8)\mu K$.
While the presence of such peak, compatible with inflationary scenario, was
already suggested by previous measurements \cite{b97}, the absence of
secondary peaks after $\ell \ge 300$ with a flat spectrum with an amplitude
of $\sim 40 \mu K$ up to $\ell \sim 625$ was a new and unexpected result.
This result obtained then an impressive confirmation by the MAXIMA-1
\cite{Maxima} experiment up to $\ell \sim 800$.

As already pointed out \cite{crisis}, the values in Eq. (\ref{standard})
for $\Omega_B h^2$, though in the correct order of magnitude, are however
somehow smaller than the baryon fraction which more easily fit the CMBR
data. In fact, the lack of observation of a secondary peak in the
anisotropy power spectrum at small scales may be a signal in favour of a
larger $\Omega_B h^2 \sim 0.03$, since increasing the baryon fraction
enhances the odd peaks only.

It has been stressed \cite{Emmp2} that a simple way to improve the
agreement of observed nuclide abundances with $\Omega_B h^2 \geq 0.02$ is
to assume non vanishing neutrino chemical potentials at the BBN epoch, a
scenario already extensively studied in the past \cite{KangSteigman}. The
effect of neutrino chemical potentials $\mu_\alpha$, with $\alpha$ the
neutrino specie, is twofold. A non-vanishing
$\xi_\alpha=\mu_{\nu_\alpha}/T_\nu$, contribute to $N_\nu$ implying a
larger expansion rate of the universe with respect to the non-degenerate
scenario, and a higher value for the neutron to proton density ratio at the
freeze-out. Furthermore, a positive value for $\xi_e$ means a larger 
number of $\nu_e$ with respect to $\bar{\nu}_e$, thus enhancing $n
\rightarrow p$ processes.

Increasing $N_\nu$ also weakly affects the CMBR anisotropy spectrum  in two
ways. The growth of perturbations inside the horizon is in fact lowered,
resulting in a decay of the gravitational potential and hence in an
increase of the anisotropy near the first peak. Moreover, the size of
horizon and sound horizon at the last scattering surface is changed, and
this, with additional effects in the damping, varies the amplitude and
position of the other peaks.

To test the degenerate BBN scenario we performed a likelihood analysis of
the data. First, to constrain the values of the parameter set $(\xi_e$,
$N_\nu$, $\Omega_B h^2)$ from the data on $^4He$, $D$ and $^7Li$ we define
a {\it total likelihood function}, ${\cal L}_{Nucl}(N_\nu, \Omega_B h^2,
\xi_e)$ \cite{Emmp2,Emmmp}. Since the effect of a positive $\xi_e$ can be
compensated by larger $N_\nu$, we have chosen to constrain this parameter
to be $N_\nu <16$, well outside the $95 \%$ upper limit on $N_\nu$ from the
BOOMERanG and MAXIMA-1 data (see below). The other two parameters are
chosen in the following ranges, $-1\leq\xi_e\leq1$ and $0.004\leq\Omega_B
h^2\leq0.110$.

\begin{figure}[t]
\vspace{9pt}
\epsfig{file=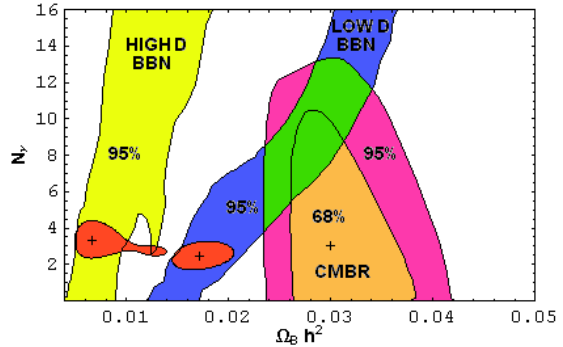,width=.47\textwidth}
\caption{The 95$\%$ C.L. contours in the $\Omega_B h^2-N_\nu$ plane
compatible with degenerate BBN (large bands) and standard BBN (small
regions) are plotted for both high $D$ (left) and low $D$ (right). The same
contours for 68$\%$ and 95$\%$ C.L. from BOOMERanG and MAXIMA-1 CMBR data
are also reported.}
\end{figure}

In the figure we summarize the main result of our analysis. In the
$\Omega_B h^2-N_\nu$ plane we show the 95$\%$ C.L. likelihood regions for
both the high and low $D$ measurements, as well as the analogous contours
for standard BBN, obtained running our code with $\xi_e=0$. In the same
plot we show the 68 and 95 $\%$ C.L. regions obtained by CMBR data.

We observe that the standard BBN, $\xi_e=0$, and CMBR data analysis lead to
quite different values for $\Omega_B h^2$. This can be clearly seen from
the reported 95$\%$ results, but we have verified that the 99$\%$ C.L.
contour for high $D$ has no overlap with the region picked up by BOOMERanG
and MAXIMA-1 data, and a very marginal one for low $D$.

For the degenerate scenario, increasing $N_\nu$, the allowed intervals for
$\Omega_B h^2$ shift towards larger values. However the high $D$ values
require a baryon content of the universe energy density which is still too
low, $\Omega_B h^2 \leq 0.018$, to be in agreement with CMBR results. A
large overlap is instead obtained for the low $D$ case, whose preferred
$\Omega_B h^2$ span the range $0.012 \leq \Omega_B h^2 \leq 0.036$. As
expected, a larger $N_\nu$ helps in improving the agreement with the high
CMBR $\Omega_Bh^2$ value, but is important to stress that a large value for
$N_\nu$ is not preferred by the CMBR data alone, being, in this case, the
best fit $N_\nu \sim 3$. If we only consider the $95 \%$ overlap region we
get the following conservative bounds\footnote{The result $N_\nu \leq 13$ 
has also been obtained in \cite{hann} using BOOMERanG data only.}:
\be
4 \leq N_\nu \leq 13~~~,~~~~~~~ 0.024 \leq \Omega_B h^2 \leq 0.034.
\ee
In this region $\xi_e$ varies in the range $0.07 \leq \xi_e \leq 0.43$. As
we said, values $N_\nu \geq 3$, as suggested from our analysis, can be
either due to weak interacting neutrino degeneracy, or rather to other
unknown relativistic degrees of freedom.

\end{document}